\documentclass[letter,traditabstract]{aa}
\usepackage{amsmath}
\usepackage{natbib,graphicx,epsfig}
\usepackage{txfonts}
\usepackage{xcolor}

\newcommand{\Msun}{M_{\odot}}
\newcommand{\Rsun}{R_{\odot}}

\begin{document}

\title{Averting the magnetic braking catastrophe on small scales: disk formation due to Ohmic dissipation}
\titlerunning{Disk formation due to Ohmic dissipation}

\author{Wolf B. Dapp and Shantanu Basu } 

\institute{Department of Physics \& Astronomy, The University of Western Ontario, 1151 Richmond St., London, ON, N6A 3K7, Canada;\\ 
email: \tt{wdapp@uwo.ca, basu@uwo.ca} }

\date{Received 7 September 2010 / Accepted 1 October 2010}

\abstract
{We perform axisymmetric resistive MHD calculations that demonstrate that centrifugal disks can indeed form around Class 0 objects despite magnetic braking. We follow the evolution of a prestellar core all the way to near-stellar densities and stellar radii. Under flux-freezing, the core is braked and disk formation is inhibited, while Ohmic dissipation renders magnetic braking ineffective within the first core. In agreement with observations that do not show evidence for large disks around Class 0 objects, the resultant disk forms in close proximity to the second core and has a radius of only $\approx 10~\Rsun$ early on. Disk formation does not require enhanced resistivity. We speculate that the disks can grow to the sizes observed around Class II stars over time under the influence of both Ohmic dissipation and ambipolar diffusion, as well as internal angular momentum redistribution.}
\keywords{magnetohydrodynamics (MHD) -- protoplanetary disks -- stars: formation -- stars: magnetic field}

\maketitle

\section{Introduction}

Understanding how protostellar and protoplanetary disks form is of fundamental importance to theories of star- and planet formation. Observations show their ubiquity around Class II objects \citep[e.g.,][]{AndrewsWilliams2005}. In recent years, doubt was cast on their accepted formation mechanism, when it was shown that for flux freezing \textit{magnetic braking} is so effective in removing angular momentum from the parent core that large-scale ($\approx 10^{2}~\mathrm{AU}$) disks are suppressed entirely \citep{AllenLiShu2003,MellonLi2008,HennebelleFromang2008}. This scenario held true even when a simplified version of ambipolar diffusion \citep{MellonLi2009} was included in the model, and has been referred to as the \textit{magnetic braking catastrophe}. Recently, \citet{HennebelleCiardi2009} demonstrated that inclination effects can modify the efficiency of magnetic braking, but a supercritical mass-to-flux ratio by a factor $>3-5$ (i.e., a weak magnetic field) was still required to form a large-scale disk. \citet{DuffinPudritz2009} performed three-dimensional simulations with ambipolar diffusion, but only resolved the first core, and did not find Keplerian motion.

Runaway collapse of a prestellar core can effectively trap the magnetic flux in the prestellar phase \citep[e.g.,][]{BasuMouschovias1994}. If the evolution continued to proceed under flux-freezing, a big magnetic flux problem would remain, since the emerging star would hold $10^{3}-10^{5}$ times more magnetic flux than observed in T Tauri stars. At densities $\lesssim 10^{12}~\mathrm{cm}^{-3}$, ambipolar diffusion causes flux leakage, while at even higher densities, matter decouples entirely from the magnetic field, and \textit{Ohmic dissipation} becomes dominant \citep[e.g.,][]{NakanoEtAl2002}. Both effects are revitalized after the formation of a central star \citep{LiMcKee1996, ContopoulosEtAl1998}. Recently, \citet{KrasnopolskyEtAl2010} have shown that for an isothermal core without self-gravity, only an `anomalous' resistivity---a factor of $100$ larger than the canonical level---allows disks of size $10^{2}~\mathrm{AU}$ to form during the Class 0 phase. However, their simulations are dominated by numerical reconnection events that make precise statements about the efficacy of magnetic braking difficult.

Currently, there is no evidence for the presence of centrifugal disks larger than $\approx 50~\mathrm{AU}$ around Class 0 or Class I objects \citep[e.g.,][]{MauryEtAl2010}.
However, there are outflows observed even at these early ages. It is therefore reasonable to assume that disks form at a small scale and only subsequently grow to the larger sizes observed in the Class II phase. We demonstrate the first part explicitly by using a canonical level of Ohmic dissipation alone, and speculate that the combined effects of ambipolar diffusion and Ohmic dissipation will allow for the second part. Additionally, an initially small disk could expand significantly if angular momentum transport is regulated by internal processes \citep[e.g.,][]{Basu1998,VorobyovBasu2007}. 

\citet{MachidaEtAl2007} performed three-dimensional simulations of resistive MHD on a nested grid, following the evolution to stellar densities, but were only able to integrate until a few days after stellar core formation. We extend their calculations in a dimensionally-simplified model in order to simultaneously address the magnetic flux problem, integrate further in time, and study the formation of a centrifugal disk. We show that catastrophic magnetic braking can be avoided, and that a small disk forms in a very early phase of evolution. 

\section{Method}\label{sec:Method}

We solve the normalized MHD equations in axisymmetric thin-disk geometry \citep[see][]{CiolekMouschovias1993,BasuMouschovias1994}, assuming vertical hydrostatic equilibrium in a vertical one-zone approximation. An integral method for calculating the self-gravity of an infinitesimally-thin disk is used \citep[detailed in][]{CiolekMouschovias1993}, with modifications for the finite extent and finite thickness of the flattened core. 

In our model, the magnetic field points solely in the vertical direction inside the disk, but also possesses radial and azimuthal components ($B_{r}$ and $B_{\phi}$) at the disk surfaces and above. $B_{r}$ is determined from a potential field assuming force-free and current-free conditions in the external medium. We calculate $B_{\phi}$ and implement magnetic braking using a steady-state approximation to the transport of Alfv\'en waves in the external medium, as in \citet{BasuMouschovias1994}. Owing to numerical complexity, a calibration of this method with results of three-dimensional MHD wave propagation through a stratified compressible medium has not been done to date. We modify the ideal-MHD induction equation to include Ohmic dissipation:
\begin{equation}
		\frac{\partial B_{z, \mathrm{eq}}}{\partial t} +
		 \frac{1}{r} \frac{\partial}{\partial r} \left( r B_{z, \mathrm{eq}} v _{r}\right)\\
		 =\frac{\eta}{r} \frac{\partial}{\partial r} \left( r \frac{\partial B_{z, \mathrm{eq}}}{\partial r} \right)
		 -\frac{\partial \eta}{\partial r} \frac{\partial B_{z, \mathrm{eq}}}{\partial r}. 
		\label{eq:induction}
\end{equation}%
Here, $B _{z, \mathrm{eq}}$ denotes the $z$-component of the magnetic field at the midplane of the disk, and $v _{r}$ is the radial component of the neutral velocity. 

We use the parametrization of \citet{MachidaEtAl2007} for the resistivity calculated by \citet{NakanoEtAl2002}, with a dimensionless scaling parameter $\widetilde{\eta} _{0}$ whose standard value is unity. The resistivity is then
\begin{multline}				
		\eta = \widetilde{\eta} _{0}~1.3\times 10^{18}~\left(\frac{n}{10^{12}~\mathrm{cm}^{-3}}\right)\left({\frac{T}{10~\mathrm{K}}}\right)^{1/2}\\
		\times\left[1-\tanh \left( \frac{n}{10^{15}~\mathrm{cm}^{-3}}\right)\right] ~\mathrm{cm}^{2}~\mathrm{s}^{-1},
		\label{eq:eta}
\end{multline}%
where $n$ is the volume number density, and the term in square brackets is a cutoff representing the restoration of flux freezing at high densities. The uncertainties in $\widetilde{\eta} _{0}$ hinge largely on the grain properties \citep[e.g.,][]{MachidaEtAl2007}. Different from \citet{MachidaEtAl2007}, we do not (inconsistently) pull the resistivity outside all spatial derivatives. 

For simplicity, we replace the detailed energy equation by a barotropic relation. The temperature-density relation of \citet{MasunagaInutsuka2000} is transformed into a pressure-density relation using the ideal gas law $P=nk_{\mathrm{B}}T$, where $P$ is the pressure, $k_{\mathrm{B}}$ is Boltzmann's constant, and $T$ is the temperature. We calculate the midplane pressure self-consistently, including the effects of the weight of the gas column, constant external pressure ($P_{\mathrm{ext}} = 0.1~\pi G \Sigma _{\mathrm{0}} ^{2}/2$), magnetic pressure, and the extra squeezing added by a central star (once present).

The MHD equations are solved with the method of lines \citep[e.g.,][]{Schiesser1991} using a finite volume approach on an adaptive grid with up to $1024$ radial cells in logarithmic spacing. The  smallest cell is initially $10^{-2}~\mathrm{AU}$ and as small as $0.02~\Rsun$ at the highest refinement. We use the second-order van-Leer TVD advection scheme \citep{vanLeer1977}, and calculate all derivatives to second-order accuracy on the nonuniform grid. The code will be described in detail in a forthcoming paper. 

\section{Initial conditions and normalization}\label{sec:Initial Conditions}

We assume that our initial state was reached by core contraction preferentially along magnetic field lines \citep[e.g.,][]{FiedlerMouschovias1993} and rotational flattening, and start with initial profiles for the column density and angular velocity given by
\begin{align}
	\Sigma \left( r \right)		= \frac{\Sigma _{0}}{\sqrt{1+\left(r/R\right)^{2}}},&&
	\Omega \left( r \right)		= \frac{2 \Omega _{\mathrm{c}}}{\sqrt{1+\left(r/R\right)^{2}}+1}.		
	\label{eq:initial_conds}
\end{align}%
Here, $R\approx 1,500~\mathrm{AU}$ approximately equals the Jeans length at the core's initial central density (see below). The column density profile is representative of the early stage of collapse \citep[e.g.,][]{Basu1997,DappBasu2009}, and the angular velocity profile reflects that the specific angular momentum of any parcel is proportional to the enclosed mass.

We assume an initial profile for $B _{z, \mathrm{eq}}$ in a way that the normalized mass-to-flux ratio $\mu=\Sigma / B_{z, \mathrm{eq}}~2 \pi \sqrt{G}=2$ everywhere, which is the approximate starting point of runaway collapse \citep[e.g.,][]{BasuMouschovias1994}. The radial velocity is initially zero. The initial state is not far from equilibrium, because the pressure gradient and magnetic and centrifugal forces add up to $\approx 82\%$ of the gravitational force. Our results do not depend strongly on the choice of initial state as long as gravity remains dominant.  

The initial central column density and number density are $\Sigma _{0} = 0.23~\mathrm{g}~\mathrm{cm}^{-2}$ and $n _{\mathrm{c}} = 4.4\times 10^{6}~\mathrm{cm}^{-3}$, respectively. The total mass and radius of the core are $2.5~\Msun$ and $1.2 \times 10^{4}~\mathrm{AU}$, respectively. The initial central magnetic field strength is $B_{z, \mathrm{eq}}\approx 200~\mu G$. We choose the external density in a way that $n _{\mathrm{c}} / n _{\mathrm{ext}} = 500$, (i.e., $n _{\mathrm{ext}} \approx 10^{3}~\mathrm{cm}^{-3}$), and the central angular velocity $\Omega _{\mathrm{c}}$ so that the cloud's edge rotates at a rate of $1~\mathrm{km}~\mathrm{s}^{-1}~\mathrm{pc}^{-1}$, consistent with observations of molecular cloud cores \citep{GoodmanEtAl1993,CaselliEtAl2002}. 

\section{Results}\label{sec:Results}

\subsection{Prestellar phase and formation of the second core}\label{subsec:results_prestellar}

During the prestellar phase (for number densities $n< 10^{11}~\mathrm{cm}^{-3}$) the collapse proceeds in a nearly self-similar fashion. We find that---insensitive to initial conditions---the column density is approximately $\propto r^{-1}$ for three orders of magnitude of central enhancement, which corresponds to the volume density being $\propto r^{-2}$ for a central enhancement of $\approx 10^{6}$. This profile is characteristic of a collapsing prestellar core \citep[e.g.,][]{Larson1969}. The collapse proceeds dynamically, and to a good approximation under isothermality, flux-freezing, and without significant magnetic braking \citep{BasuMouschovias1994}.

Once the density reaches $\approx 10^{11}~\mathrm{cm}^{-3}$, the central region becomes opaque and traps the energy released by the collapse, which previously could escape freely as radiation. This region heats up \citep{Larson1969, MasunagaInutsuka2000} and its thermal pressure gradient temporarily stabilizes it against further collapse. This is the \textit{first core}. Its density and temperature increase with continued accretion, while its size stays almost constant at $\approx$ a few AU, bounded by an accretion shock. The external gravitational potential of this object closely resembles that of a point mass, and an expansion wave develops and moves outward at nearly the sound speed \citep{Shu1977}. Material within this region moves at near free-fall speed. 

When the temperature in the first core reaches $\approx$ $2000~\mathrm{K}$, for $n \gtrsim 10^{15}~\mathrm{cm}^{-3}$, hydrogen molecules are collisionally dissociated. This process provides an energy sink, so that the temperature rise stagnates, and the collapse reinitiates. As the temperature rises yet further, hydrogen is ionized sufficiently that flux freezing is re-established. Collapse is then finally halted, and sufficiently high densities are reached that electron degeneracy becomes important \citep{MasunagaInutsuka2000}. A protostellar core (the \textit{second core}) forms with a radius $\approx$ a few $\Rsun$ \citep[e.g.,][]{Larson1969}. This Class 0 object initially only has a mass of a few $\times 10^{-3}~\Msun$. The gravitational potential resembles that of a point mass outside the second core, and an expansion wave once again moves outward from the accretion shock, eventually consuming the entire region of the previous first core. 

\begin{figure}
  \includegraphics[width=0.9\hsize]{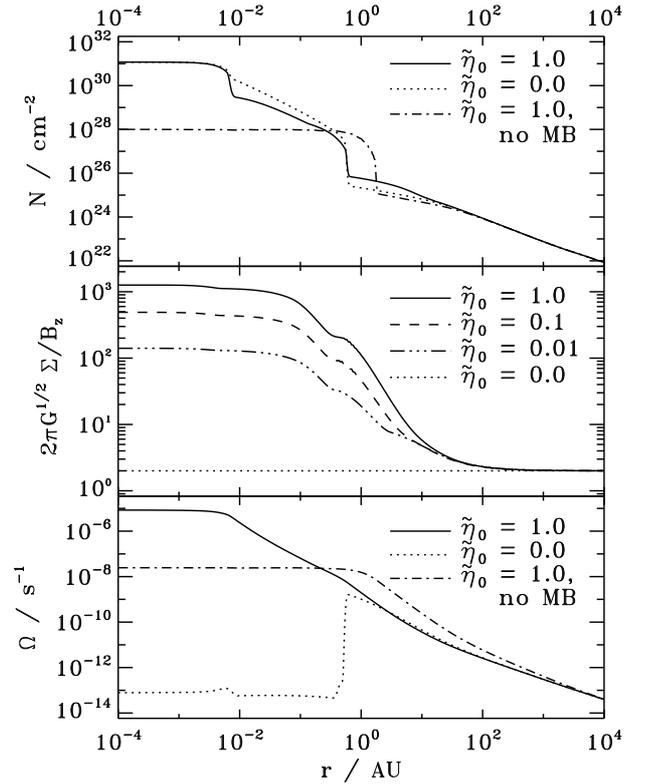}  
      \centering
      \caption{
      	Spatial profiles of various quantities after the second collapse (after $\approx 4.8 \times 10^{4}~\mathrm{yr}$).
      	\textbf{Top:} The first and second core and their accretion shocks are at radii 
      	$\approx 1~\mathrm{AU}$ and $\approx 5 \times 10^{-3}~\mathrm{AU}\approx 1~\Rsun$, respectively. Within the expansion wave outside the first core, 
      	the column density profile assumes that of free-fall collapse in the flux-freezing case ($\widetilde{\eta} _{0}=0$), and shows a 
      	magnetic wall in the resistive case. Beyond $\approx 20~\mathrm{AU}$, the prestellar infall profile remains unchanged. Without magnetic braking (dash-dotted line), the first core is larger and rotation prevents further collapse.
      	\textbf{Middle:} The mass-to-flux ratio is increased by (even weak) Ohmic dissipation	by $\gtrsim 10^{2}$. 
      	The influence is significant even well outside the boundary of the first core (at a few AU). 
      	\textbf{Bottom:} For flux-freezing, catastrophic magnetic braking spins down the first core to nearly the background rotation rate. In the resistive case (solid line), the rotation rate outside the first core is reduced only slightly compared with the case without magnetic braking (dash-dotted line).
      }
      \label{fig:sigma_omega_mu}
\end{figure}

Figure \ref{fig:sigma_omega_mu} shows the profiles of column density, mass-to-flux ratio and angular velocity shortly after the second core forms ($\approx 4.8 \times 10^{4}~\mathrm{yr}$ into the simulation). For $n \gtrsim 10^{12}~\mathrm{cm}^{-3}$, Ohmic dissipation becomes dynamically important \citep{NakanoEtAl2002}, because all charge carriers decouple from the magnetic field, and flux is dissipated. While the density in the first core increases, we find the magnetic field strength remains stagnant. A \textit{magnetic wall} \citep{LiMcKee1996,ContopoulosEtAl1998} forms at $\approx 10~\mathrm{AU}$, visible as a sharp transition in column density in the resistive model ($\widetilde{\eta} _{0} = 1$, top panel). Here, infalling neutrals within the expansion wave are temporarily slowed down by the relatively well-coupled magnetic field that is expelled from the first core with a radius $\approx 1~\mathrm{AU}$. Further inward, the neutrals resume near-free-fall motion, but with enhanced magnetic support and at a greater column density than for flux-freezing ($\widetilde{\eta} _{0} = 0$, dotted line). Under angular momentum conservation (no magnetic braking), the additional rotational support stabilizes the first core against further collapse (top panel, dash-dotted line), consistent with previous findings \citep[e.g.,][]{SaigoTomisaka2006}. 

Because of magnetic flux dissipation, the mass-to-flux ratio increases by almost three orders of magnitude in the first core region for $\widetilde{\eta} _{0} = 1$, but by almost two orders of magnitude even for $\widetilde{\eta} _{0}$ as low as $0.01$ (Fig. \ref{fig:sigma_omega_mu}, middle panel). The torque on the cloud caused by magnetic braking scales linearly with the amount of enclosed flux \citep[][]{BasuMouschovias1994}. Ohmic dissipation therefore allows spin-up to proceed, even though the rotation rate is still reduced by a factor of a few outside the first core, compared with the case without magnetic braking (Fig. \ref{fig:sigma_omega_mu}, bottom panel, dash-dotted line). In the flux-freezing case, the comparatively slow evolution of the first core allows enough time for magnetic braking to spin down the first core region, and `catastrophically' brake it (Fig. \ref{fig:sigma_omega_mu}, bottom panel, dotted line).

\subsection{Evolution after second core formation}\label{subsec:results_protostellar}

When the second core forms, the thin-disk formulation breaks down, because the object is now truly hydrostatic and spherical. Presumably, dynamo processes within the fully convective protostar will also take over, and the magnetic field will mostly decouple from that of its parent core \citep{MestelLandstreet2005}. Therefore, we switch off magnetic braking in the second core, and introduce a sink cell with a size of $3~\Rsun$, slightly larger than the second core. The processes within it are beyond the scope of our model, but are not expected to significantly influence the surroundings. This is not necessarily the case with a sink cell of size $\approx 10~\mathrm{AU}$, as is the more common approach \citep[e.g.,][]{VorobyovBasu2007,MellonLi2008,MellonLi2009}.

\begin{figure}
  \includegraphics[width=0.9\hsize]{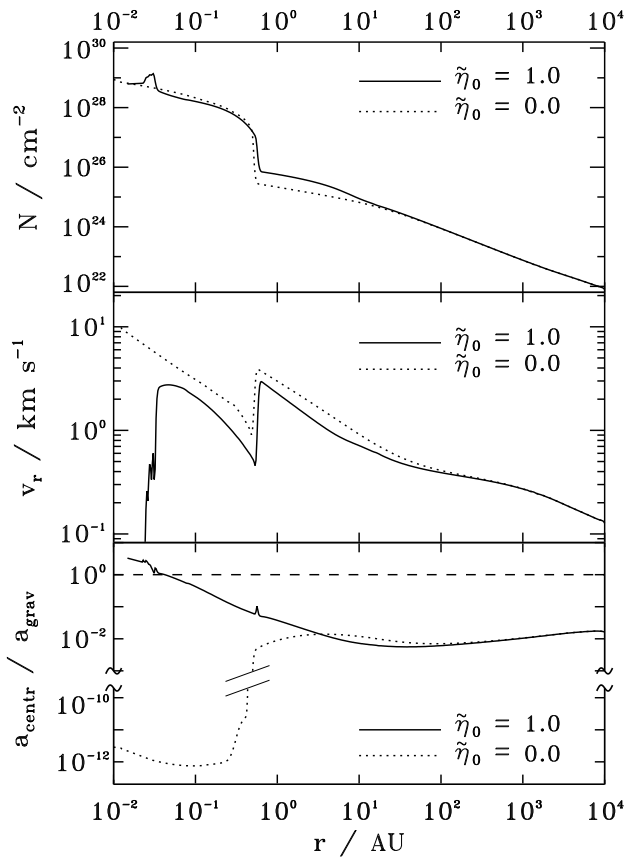}
      \centering
      \caption{
      	Spatial profiles of various quantities $\approx 1~\mathrm{yr}$ after the introduction of a sink cell 
      	of size $\approx 3~\Rsun$.
      	\textbf{Top:} The Toomre-unstable centrifugally-supported disk breaks up into a ring. 
      	\textbf{Middle:} Infall is halted after the formation of a centrifugal disk at 
      	$\approx 5\times 10^{-2}~\mathrm{AU}\approx 10~\Rsun$ in the resistive case ($\widetilde{\eta} _{0}=1$),
      	while for flux-freezing ($\widetilde{\eta} _{0}=0$), infall continues.       	
      	\textbf{Bottom:} Ratio between centrifugal and gravitational accelerations. The dashed line indicates 
      	rotational balance, achieved within $\approx 10~\Rsun$ 
      	with Ohmic dissipation. For flux-freezing, rotational support is negligible in the first core region 
      	owing to the magnetic braking catastrophe.
      }
      \label{fig:sigma_vr_acc_disk}
\end{figure}

Figure \ref{fig:sigma_vr_acc_disk} shows the profiles of column density, infall velocity, and the ratio of centrifugal to gravitational acceleration about a year after the introduction of the sink cell. Centrifugal balance is achieved in a small region ($\approx 10~\Rsun$) close to the center (bottom panel) in the resistive model. This is a necessary and sufficient condition for the formation of a centrifugally-supported disk. At the same time all infall is halted there and the radial velocity plummets (middle panel). After a few years of evolution, a Toomre instability develops, and the rotationally-supported structure breaks up into a ring (top panel). At this point, we stop the simulation, because more physics would be required to follow the further evolution of the disk. Our model allows a clear distinction between a magnetic pseudo-disk, a flattened (disk-like) prestellar core, and a centrifugal (nearly Keplerian) disk. This distinction is not clear in profiles from three-dimensional simulations \citep{MachidaEtAl2007,DuffinPudritz2009}.

\begin{figure}
	\begin{minipage}[b]{0.46\hsize}
		\centering
		\includegraphics[scale=0.57]{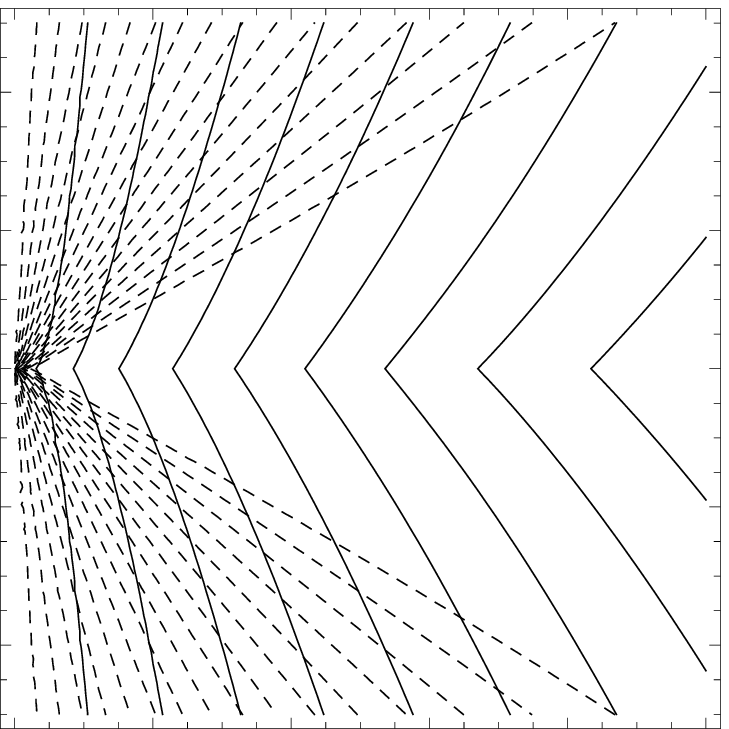}		
	\end{minipage}
	\hspace{0.5cm}
	\begin{minipage}[b]{0.46\hsize}
		\centering
		\includegraphics[scale=0.45]{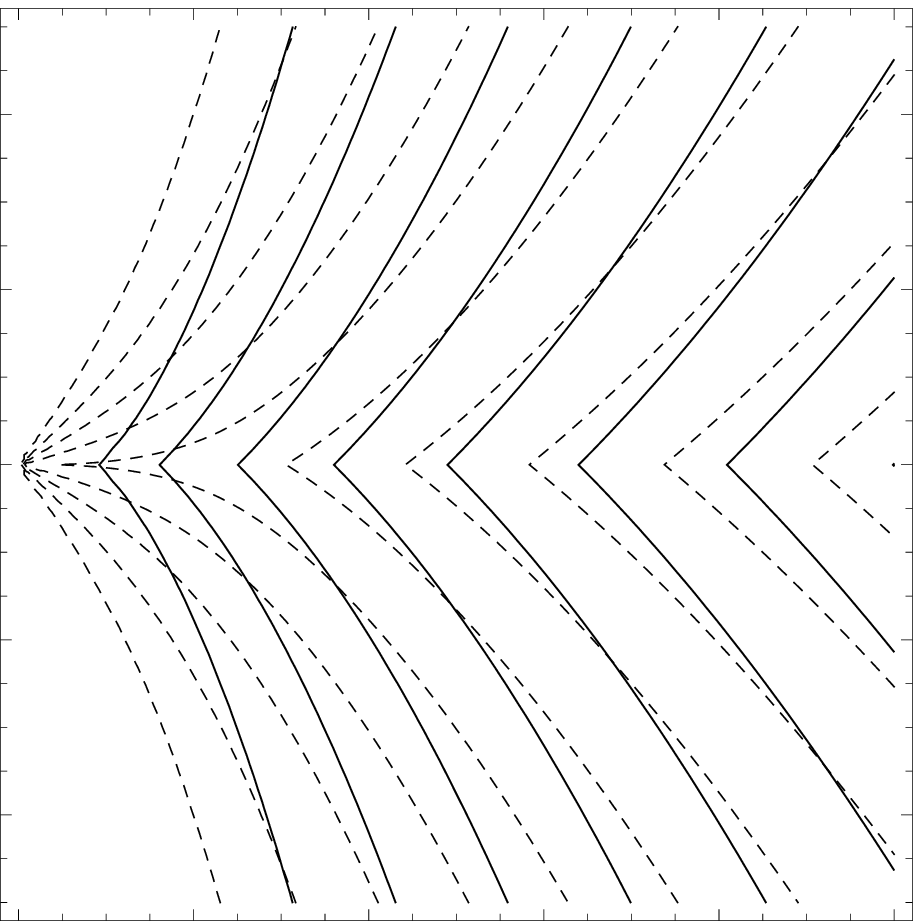}		
	\end{minipage}		
	\caption{Magnetic field lines. The box on the left has dimensions $10~\mathrm{AU}$ on each side, while the box on the right has dimensions $100~\mathrm{AU}$. The dashed lines represent the flux-freezing model ($\widetilde{\eta} _{0} = 0$), while the solid lines show the same field lines for the resistive model ($\widetilde{\eta} _{0} = 1$). The second core has just formed and is on the left axis midplane.}		
	\label{fig:fieldlines}
\end{figure}

Figure \ref{fig:fieldlines} shows the magnetic field line topology above and below the disk on two scales ($10~\mathrm{AU}$ and $100~\mathrm{AU}$), for both flux-freezing and resistive models. They are calculated immediately after the formation of the second core, assuming force-free and current-free conditions above a thin disk \citep{MestelRay1985}. The split monopole of the $\widetilde{\eta} _{0} = 0$ model (dashed lines) is created as field lines are dragged in by the freely falling material within the expansion wave front at $\approx 20~\mathrm{AU}$. This is replaced by a much more relaxed field line structure in the resistive case (solid lines). The extreme flaring of field lines in the $\widetilde{\eta} _{0} = 0$ model is a fundamental cause of the magnetic braking catastrophe. \citet{GalliEtAl2009} presented similar field configurations resulting from a simplified model for resistive collapse. 

\section{Discussion and conclusions}\label{sec:Discussion}

We demonstrate the formation of a centrifugally-supported disk despite the presence of magnetic braking. The magnetic braking catastrophe is averted by including the canonical level of Ohmic dissipation, which removes large amounts of magnetic flux from the high-density region of the first core. In the absence of Ohmic dissipation, this region would be spun down tremendously prior to the second collapse. We emphasize that disk formation happens very shortly after the second collapse in a region very close to the central object, while it is still very small ($<10^{-2}~\Msun$). This is consistent with the observational evidence of outflows at a very young age. 

Our simulations yield $\approx 0.1-1~\mathrm{kG}$ magnetic fields, comparable to those observed in T Tauri stars \citep[e.g.,][]{JohnsKrull2007}, in a central object of mass $\approx 10^{-2}~\Msun$. This is achieved by non-ideal MHD effects reducing the field strength by $\approx 10^{3}$ compared to a flux-freezing model. Our model does not have the capability of including outflows or jets, even though those are launched very close to the stellar surface. 

There is presently no evidence for centrifugal disks $\gtrsim 50~\mathrm{AU}$ around Class 0 objects \citep[e.g.,][]{AndreEtAl2002, MauryEtAl2010}. ALMA will allow observers to improve on this, and to probe for disks down to $\approx 10~\mathrm{AU}$. We anticipate that the centrifugal disk that forms in our simulations \textit{can} grow over time into disks of size $\approx 100~\mathrm{AU}$ observed around Class II objects. 
Recent work \citep{MachidaEtAl2010} shows that magnetic braking can be cut off at late times as the envelope is accreted, and the existing disk can also grow by internal angular momentum redistribution processes \citep[e.g.,][]{VorobyovBasu2007}. Furthermore, we speculate that ambipolar diffusion \citep[e.g.,][]{KunzMouschovias2010} has the potential to dissipate enough flux outside the first core (an area not significantly affected by Ohmic dissipation) to reduce braking and to allow the disk to form there as well. We will present results of a study including both non-ideal MHD effects and grain physics in an upcoming paper. 

\section*{Acknowledgments}
The authors thank the participants of the CC2YSO conference for engaging and illuminating discussions. W.B.D. was supported by an NSERC Alexander Graham Bell Canada Graduate Scholarship, and S.B. by an NSERC Discovery Grant.


\begin{thebibliography}{}

\bibitem[\protect\citeauthoryear{Andr\'e et al.}{2002}]{AndreEtAl2002} Andr\'e, P., Bouwman, J., Belloche, A., \& Hennebelle, P. 2002, Ap\&SS, 292, 325

\bibitem[\protect\citeauthoryear{Andrews \& Williams}{2005}]{AndrewsWilliams2005} Andrews, S. M., \& Williams, J. P. 2005, ApJ, 631, 1134

\bibitem[\protect\citeauthoryear{Allen, Li, \& Shu}{2003}]{AllenLiShu2003} Allen, A., Li, Z.-Y., \& Shu, F. H. 2003, ApJ, 599, 363

\bibitem[\protect\citeauthoryear{Basu}{1997}]{Basu1997} Basu, S. 1997, ApJ, 485, 240

\bibitem[\protect\citeauthoryear{Basu}{1998}]{Basu1998} Basu, S. 1998, ApJ, 509, 229

\bibitem[\protect\citeauthoryear{Basu \& Mouschovias}{1994}]{BasuMouschovias1994} Basu, S., \& Mouschovias, T. Ch. 1994, ApJ, 432, 720

\bibitem[\protect\citeauthoryear{Caselli et al.}{2002}]{CaselliEtAl2002} Caselli, P., Benson, P., Myers, P. C., \& Tafalla, M. 2002, ApJ, 572, 238

\bibitem[\protect\citeauthoryear{Ciolek \& Mouschovias}{1993}]{CiolekMouschovias1993} Ciolek, G. E., \& Mouschovias, T. Ch. 1993, ApJ, 418, 774

\bibitem[\protect\citeauthoryear{Contopoulos et al.}{1998}]{ContopoulosEtAl1998} Contopoulos, I., Ciolek, G. E., \& K\"onigl, A. 1998, ApJ, 504, 247

\bibitem[\protect\citeauthoryear{Dapp \& Basu}{2009}]{DappBasu2009} Dapp, W. B., \& Basu, S. 2009, MNRAS, 395, 1092

\bibitem[\protect\citeauthoryear{Duffin \& Pudritz}{2009}]{DuffinPudritz2009} Duffin, D. F., \& Pudritz, R. E. 2009, ApJ, 706, L46

\bibitem[\protect\citeauthoryear{Fiedler \& Mouschovias}{1993}]{FiedlerMouschovias1993} Fiedler, R. A., \& Mouschovias, T. Ch. 1993, ApJ, 415, 640

\bibitem[\protect\citeauthoryear{Galli et al.}{2009}]{GalliEtAl2009} Galli, D., Cai, M., Lizano, S., \& Shu, F. H. 2009, RevMexAA, 36, 143

\bibitem[\protect\citeauthoryear{Goodman et al.}{1993}]{GoodmanEtAl1993} Goodman, A. A., Benson, P. J., Fuller, G. A., \& Myers, P. C. 1993, ApJ, 406, 528
 
\bibitem[\protect\citeauthoryear{Hennebelle \& Ciardi}{2009}]{HennebelleCiardi2009} Hennebelle, P., \& Ciardi, A. 2009, A\&A, 506, L29

\bibitem[\protect\citeauthoryear{Hennebelle \& Fromang}{2008}]{HennebelleFromang2008} Hennebelle, P., \& Fromang, S. 2008, A\&A, 477, 9

\bibitem[\protect\citeauthoryear{Johns-Krull}{2007}]{JohnsKrull2007} Johns-Krull, C. M. 2007, ApJ, 664, 975
	
\bibitem[\protect\citeauthoryear{Krasnopolsky et al.}{2010}]{KrasnopolskyEtAl2010} Krasnopolsky, R., Li, Z.-Y., \& Shang, H. 2010, ApJ, 716, 1541

\bibitem[\protect\citeauthoryear{Kunz \& Mouschovias}{2010}]{KunzMouschovias2010} Kunz, M. W., \& Mouschovias, T. Ch. 2010, MNRAS, 408, 322
 
\bibitem[\protect\citeauthoryear{Larson}{1969}]{Larson1969} Larson, R. B. 1969, MNRAS, 145, 271

\bibitem[\protect\citeauthoryear{van Leer}{1977}]{vanLeer1977} van Leer, B. 1977, JCP, 23, 276

\bibitem[\protect\citeauthoryear{Li \& McKee}{1996}]{LiMcKee1996} Li, Z.-Y., \& McKee, C. F. 1996, ApJ, 464, 373

\bibitem[\protect\citeauthoryear{Machida et al.}{2007}]{MachidaEtAl2007} Machida, M. N., Inutsuka, S.-i., \& Matsumoto, T. 2007, ApJ, 670, 1198

\bibitem[\protect\citeauthoryear{Machida et al.}{2010}]{MachidaEtAl2010} Machida, M. N., Inutsuka, S.-i., \& Matsumoto, T. arXiv:1009.2140v1

\bibitem[\protect\citeauthoryear{Masunaga \& Inutsuka}{2000}]{MasunagaInutsuka2000} Masunaga, H., \& Inutsuka, S.-i. 2000, ApJ, 531, 350

\bibitem[\protect\citeauthoryear{Maury et al.}{2010}]{MauryEtAl2010} Maury, A. J., Andr\'e, P., Hennebelle, P., et al. 2010, A\&A, 512, 40

\bibitem[\protect\citeauthoryear{Mellon \& Li}{2008}]{MellonLi2008} Mellon, R. R., \& Li, Z.-Y. 2008, ApJ, 681, 1356

\bibitem[\protect\citeauthoryear{Mellon \& Li}{2009}]{MellonLi2009} Mellon, R. R., \& Li, Z.-Y. 2009, ApJ, 698, 922

\bibitem[\protect\citeauthoryear{Mestel \& Landstreet}{2005}]{MestelLandstreet2005} Mestel, L., \& Landstreet, J. D. 2005, in Lecture Notes in Physics vol. 664, Cosmic Magnetic Fields, ed. R. Wielebinski, R. Beck, 183

\bibitem[\protect\citeauthoryear{Mestel \& Ray}{1985}]{MestelRay1985} Mestel, L., \& Ray, T. P. 1985, MNRAS, 212, 275

\bibitem[\protect\citeauthoryear{Nakano, Nishi, \& Umebayashi}{Nakano et al.}{2002}]{NakanoEtAl2002} Nakano, T., Nishi, R., \& Umebayashi, T. 2002, ApJ, 573, 199

\bibitem[\protect\citeauthoryear{Saigo \& Tomisaka}{2006}]{SaigoTomisaka2006} Saigo, K., \& Tomisaka, K. 2006, ApJ, 645, 381

\bibitem[\protect\citeauthoryear{Schiesser}{1991}]{Schiesser1991} Schiesser, W. E. 1991, The Numerical Method of Lines: Method of Integration of Partial Differential Equations (San Diego: Academic)

\bibitem[\protect\citeauthoryear{Shu}{1977}]{Shu1977} Shu, F. H. 1977, ApJ, 214, 488

\bibitem[\protect\citeauthoryear{Vorobyov \& Basu}{2007}]{VorobyovBasu2007} Vorobyov, E. I., \& Basu, S. 2007, MNRAS, 381, 1009

\end{thebibliography}
\end{document}